\documentclass{aa}
\usepackage{txfonts}

\usepackage{newtxtext,newtxmath,upgreek}

\usepackage{graphicx}	

\usepackage{xcolor}
\usepackage[breaklinks=true,colorlinks=true,allcolors=blue]{hyperref}

\usepackage{amsmath}	
\usepackage{amssymb}	
\usepackage{natbib, booktabs, multirow, tabularx} 
\usepackage{hhline}

\usepackage{threeparttable}

\usepackage{float}
\restylefloat{figure}

\begin{document}

\title{A search for radio emission from double-neutron star merger GW190425 using Apertif}

\author{
     Oliver~M.~Boersma      \inst{\ref{uva} \and \ref{astron}}
\and Joeri~van~Leeuwen    \inst{\ref{astron} \and \ref{uva}}
\and Elizabeth~A.~K.~Adams \inst{\ref{astron} \and \ref{kapteyn}}
\and Björn~Adebahr       \inst{\ref{airub}}
\and Alexander~Kutkin     \inst{\ref{astron} \and \ref{lebedev}}
\and Tom~Oosterloo        \inst{\ref{astron} \and \ref{kapteyn}}
\and W.~J.~G.~de~Blok     \inst{\ref{astron} \and \ref{cpt} \and \ref{kapteyn}}
\and R.~van~den~Brink     \inst{\ref{astron} \and \ref{tricas}}
\and A.~H.~W.~M.~Coolen   \inst{\ref{astron}}
\and L.~Connor            \inst{\ref{uva} \and \ref{caltech}}
\and S.~Damstra           \inst{\ref{astron}}
\and H.~Dénes           \inst{\ref{astron}}
\and K.~M.~Hess      \inst{\ref{astron} \and \ref{kapteyn}}   
\and J.~M.~van~der~Hulst  \inst{\ref{kapteyn}}  
\and B.~Hut               \inst{\ref{astron}}
\and M.~Ivashina          \inst{\ref{chalmers}}
\and G.~M.~Loose      \inst{\ref{astron}}  
\and D.~M.~Lucero         \inst{\ref{virginiatech}}
\and Y.~Maan          \inst{\ref{astron}}  
\and {\'A}.~Mika             \inst{\ref{astron}}
\and V.~A.~Moss           \inst{\ref{csiro} \and \ref{sydney} \and \ref{astron}}
\and H.~Mulder            \inst{\ref{astron}}
\and L.~C.~Oostrum        \inst{\ref{astron} \and \ref{uva}}
\and M.~Ruiter            \inst{\ref{astron}}
\and D.~van~der~Schuur    \inst{\ref{astron}}
\and R.~Smits             \inst{\ref{astron}}
\and N.~J.~Vermaas        \inst{\ref{astron}}
\and D.~Vohl            \inst{\ref{astron}}  
\and J.~Ziemke            \inst{\ref{astron} \and \ref{rugcit}}
}

\institute{
Anton Pannekoek Institute, University of Amsterdam, Postbus 94249, 1090 GE Amsterdam, The Netherlands\label{uva}
  \and
ASTRON, the Netherlands Institute for Radio Astronomy, Oude Hoogeveensedijk 4,7991 PD Dwingeloo, The Netherlands\label{astron}
  \and
Kapteyn Astronomical Institute, PO Box 800, 9700 AV Groningen, The Netherlands\label{kapteyn}
  \and
Astronomisches Institut der Ruhr-Universit{\"a}t Bochum (AIRUB), Universit{\"a}tsstrasse 150, 44780 Bochum, Germany\label{airub}
  \and
Astro Space Center of Lebedev Physical Institute, Profsoyuznaya Str. 84/32, 117997 Moscow, Russia\label{lebedev}
  \and
Dept.\ of Astronomy, Univ.\ of Cape Town, Private Bag X3, Rondebosch 7701, South Africa\label{cpt}
  \and
Tricas Industrial Design \& Engineering, Zwolle, The Netherlands\label{tricas}
  \and
Cahill Center for Astronomy, California Institute of Technology, Pasadena, CA, USA\label{caltech}
  \and
Dept.\ of Electrical Engineering, Chalmers University of Technology, Gothenburg, Sweden\label{chalmers}
  \and
Department of Physics, Virginia Polytechnic Institute and State University, 50 West Campus Drive, Blacksburg, VA 24061, USA\label{virginiatech}
  \and
CSIRO Astronomy and Space Science, Australia Telescope National Facility, PO Box 76, Epping NSW 1710, Australia\label{csiro}
  \and
Sydney Institute for Astronomy, School of Physics, University of Sydney, Sydney, New South Wales 2006, Australia\label{sydney}
  \and
Rijksuniversiteit Groningen Center for Information Technology, P.O. Box 11044, 9700 CA Groningen, the Netherlands\label{rugcit}
}

\abstract
{Detection of the electromagnetic emission from coalescing 
binary neutron stars (BNS) is important for understanding 
the merger and afterglow.} 
{We present a search for a radio counterpart to the gravitational-wave source GW190425, a BNS merger, using Apertif on the Westerbork Synthesis Radio Telescope (WSRT).} 
{We observe a field of high probability in the associated localisation region 
for 3 epochs at $\Delta T =$ 68, 90, 109 d post merger. We identify all sources that exhibit flux variations 
consistent with the expected afterglow emission of GW190425. We also look for possible transients. These are sources which are only present in one epoch. In addition, we quantify our ability to search for radio afterglows in fourth and future observing runs of the gravitational-wave detector network using Monte Carlo simulations.} 
{We found 25 afterglow candidates based on their variability. None of these could be associated with a possible host galaxy at the luminosity distance of GW190425. We also found 55 transient afterglow candidates that were only detected in one epoch. All turned out to be image artefacts. In the fourth observing run, we predict that up to three afterglows will be detectable by Apertif.}
{While we did not find a source related to the afterglow emission of GW190425, the search validates our methods for future searches of radio afterglows.}


\maketitle
\section{Introduction}
\begin{figure}[ht]
    \centering
    \includegraphics[width=0.45\textwidth]{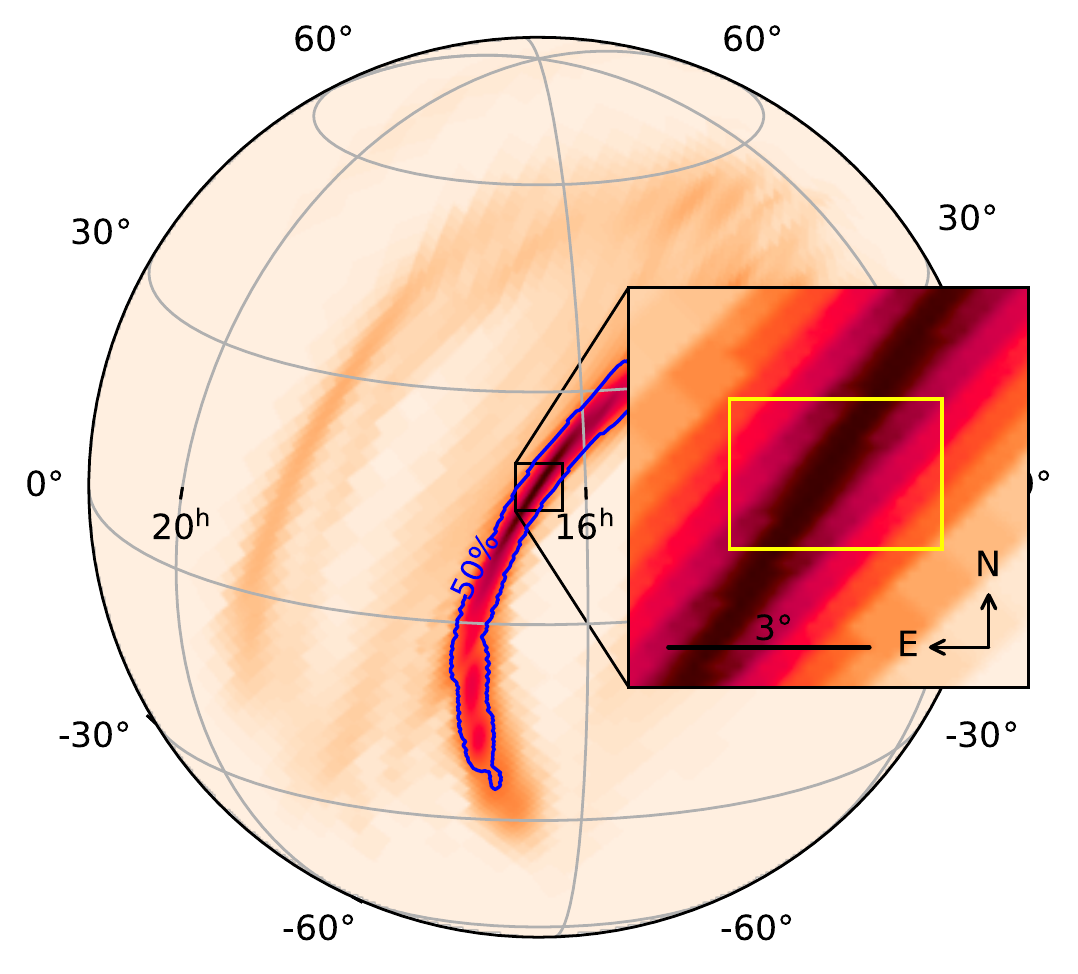}
    \caption{A part of the LIGO/Virgo localisation skymap on the northern hemisphere for GW190425. The full skymap has a 90\% credible sky area of 7461 $\mathrm{deg}^2$ which reduces to 1378 $\mathrm{deg}^2$ at the 50\% level. The blue contour encompasses the 50\% credible sky area on this part of the sky. The inset shows the approximate Apertif field-of-view in yellow.}
    \label{fig:skymap}
\end{figure}
The Advanced LIGO detector network
started its first observing run in 2015  
and was joined by  
 the Advanced Virgo detector for the second run ~\citep{aasi_advanced_2015,acernese_advanced_2014}.
 These runs yielded ten detections of binary black hole (BBH) mergers and one binary neutron star (BNS) merger~\citep{ligo_scientific_collaboration_and_virgo_collaboration_gwtc-1_2019}. The successful detection of electromagnetic (EM) emission associated with the sole BNS merger, GW170817~\citep{ligo_scientific_collaboration_and_virgo_collaboration_gw170817_2017}, included the dynamical ejecta of a kilonova (KN; see e.g., \citealt{chornock_electromagnetic_2017,coulter_swope_2017}
), 
the short gamma-ray burst  GRB170817A~\citep[e.g.,][]{abbott_gravitational_2017}
, and the afterglow of the jet interacting with the interstellar environment~\citep[e.g.,][]{alexander_electromagnetic_2017,haggard_deep_2017,hallinan_radio_2017}. Together these studies brought forth an exciting new chapter in multi-messenger astronomy. 
\begin{figure*}[ht]
\centering
\includegraphics[width=0.8\textwidth]{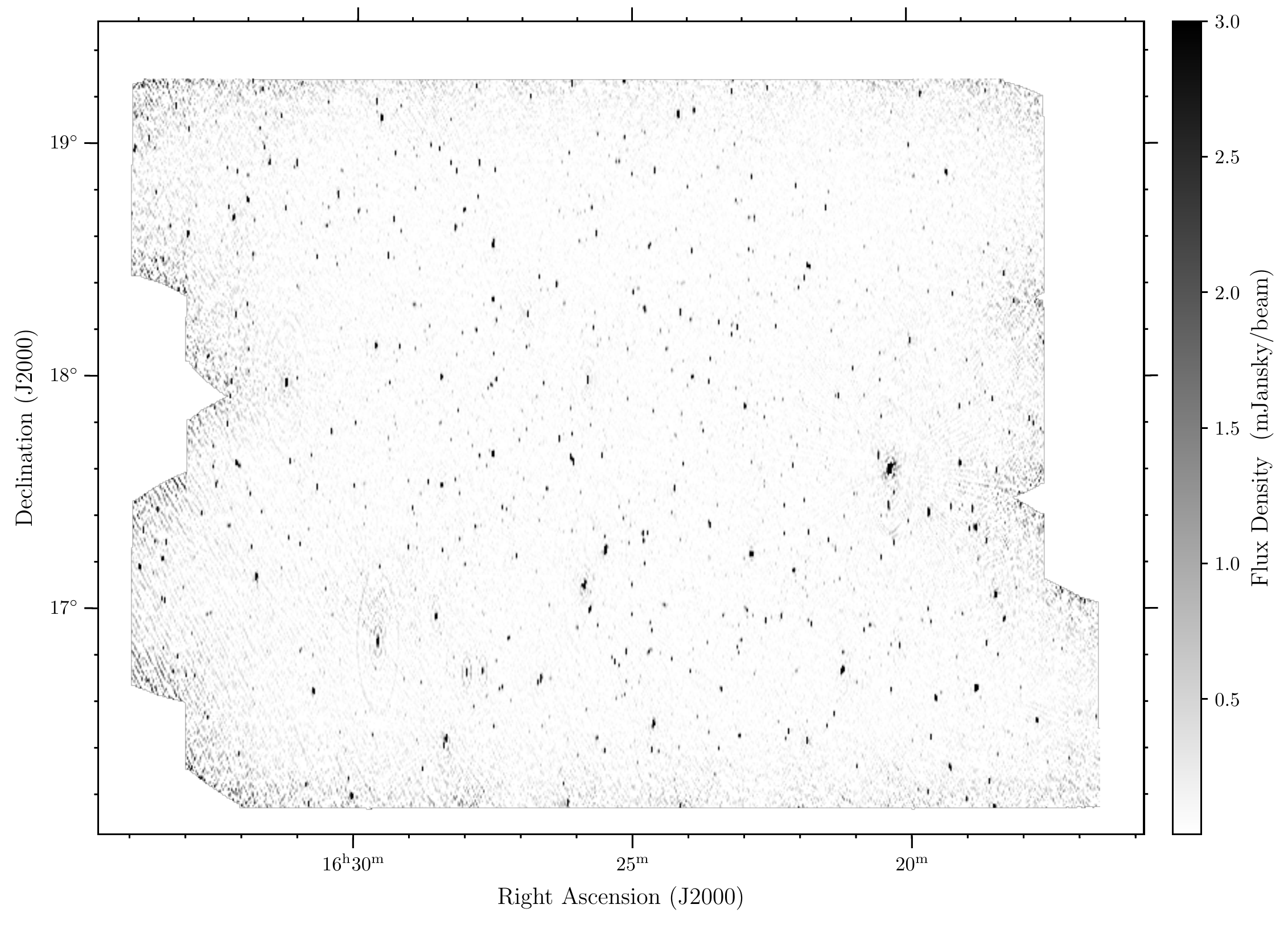}
\caption{Apertif mosaic, centered on 16:25:12.0, 17:47:24.0, covering a region of high probability of in the localisation skymap of GW190425. The mosaic, our third epoch,  was observed 109 days post merger.}
\label{fig:mosaic}
\end{figure*}
\begin{table*}[h]
\begin{tabular}{llllllll}
\hhline{========}
Epoch & ObsID     & Start (UTC)         & Int. Time (h:m:s) & $\Delta$T (day) & FOV ($\mathrm{deg}^2$) & Sensitivity & Beam Size                        \\ 
&&&&&& (best, $\mu \mathrm{Jy}$)&  \\\hline
1     & 190701042 & 2019-07-01 15:21:17 & 11:58:01          & 68              &           9.3     & 70 &        90" $\times$ 23" \\
2     & 190724131 & 2019-07-24 13:51:34 & 11:58:01          & 90              &           8.7     & 60 &       50" $\times$ 12" \\
3     & 190812081 & 2019-08-12 12:37:00 & 12:00:01          & 109             &           9.3     & 50 &    48" $\times$ 11" \\ \hline
\end{tabular}
\caption{Overview of the three epochs of observations with Apertif.}
\label{tab:obssum}
\end{table*}
In the first half of the third observing run (O3), the Advanced LIGO and Virgo detector network
detected 39 candidate compact binary coalescences~\citep{abbott_gwtc-2_2020}. The first detection of a BNS candidate in this run, LIGO/Virgo S190425z, later confirmed as GW190425~\citep{abbott_gw190425_2020}, occurred on April 25 2019.
It was observed solely by the Advanced LIGO Livingston detector at a distance of $159^{+69}_{-72}$ Mpc. The Advanced Virgo detector was also operational but did not detect the merger. The localisation capabilities of just a single detector are poor and, as such, the LAL Inference~\citep{veitch_parameter_2015} skymap (\texttt{LALInference.fits.gz}) for this source, shown partly in Fig.~\ref{fig:skymap}, has a 90\% credible localisation region on the sky of 7461 $\mathrm{deg}^2$. Nonetheless, follow-up campaigns using optical and/or infrared facilities were performed directly after the merger. 
Campaigns such as GROWTH~\citep{coughlin_growth_2019} and GRANDMA~\citep{gendre_ligovirgo_2019} aimed to find coincident EM emission from a KN or  short gamma-ray burst (SGRB) counterpart. While these efforts covered a large region of the probability map, the searches did not lead to identifying a viable source of the GW emission.

In the absence of an optical or infrared counterpart, radio emission may provide the only way to localise the source at later times~\citep{hotokezaka_radio_2016}. In order to maximise the probability of detecting coincident radio emission, a relatively large field-of-view and high sensitivity are necessary.   
Here, we present a follow-up of GW190425 with the new Apertif phased array feeds \citep[PAFs;][]{oosterloo_latest_2010,adams_radio_2019} on the Westerbork Synthesis Radio Telescope (WSRT). 
While the fields-of-view of these PAFs, at 9.5 $\mathrm{deg}^2$ \citep{artsso20}, are one of the largest in the world,
they are still dwarfed by the error region.
As the detection of an afterglow has the potential to significantly further our understanding of such mergers,
we performed the search against these odds. 
In Section \ref{sec:reduction} we discuss our observations and the data reduction methods. We examine the flux consistency between our three observations in Section \ref{sec:fluxcons}. In Section \ref{sec:transientsearch} we describe the search for radio transients in our observations. We make predictions for the amount of radio afterglows Apertif will detect in the fourth and future gravitational-wave detector network observing runs in Section \ref{sec:obsprosp}, and conclude in Section \ref{sec:conclusion}.

\section{Observations \& Data Reduction}\label{sec:reduction}
\begin{figure*}[ht]
    \centering
    \includegraphics[width=0.6\textwidth]{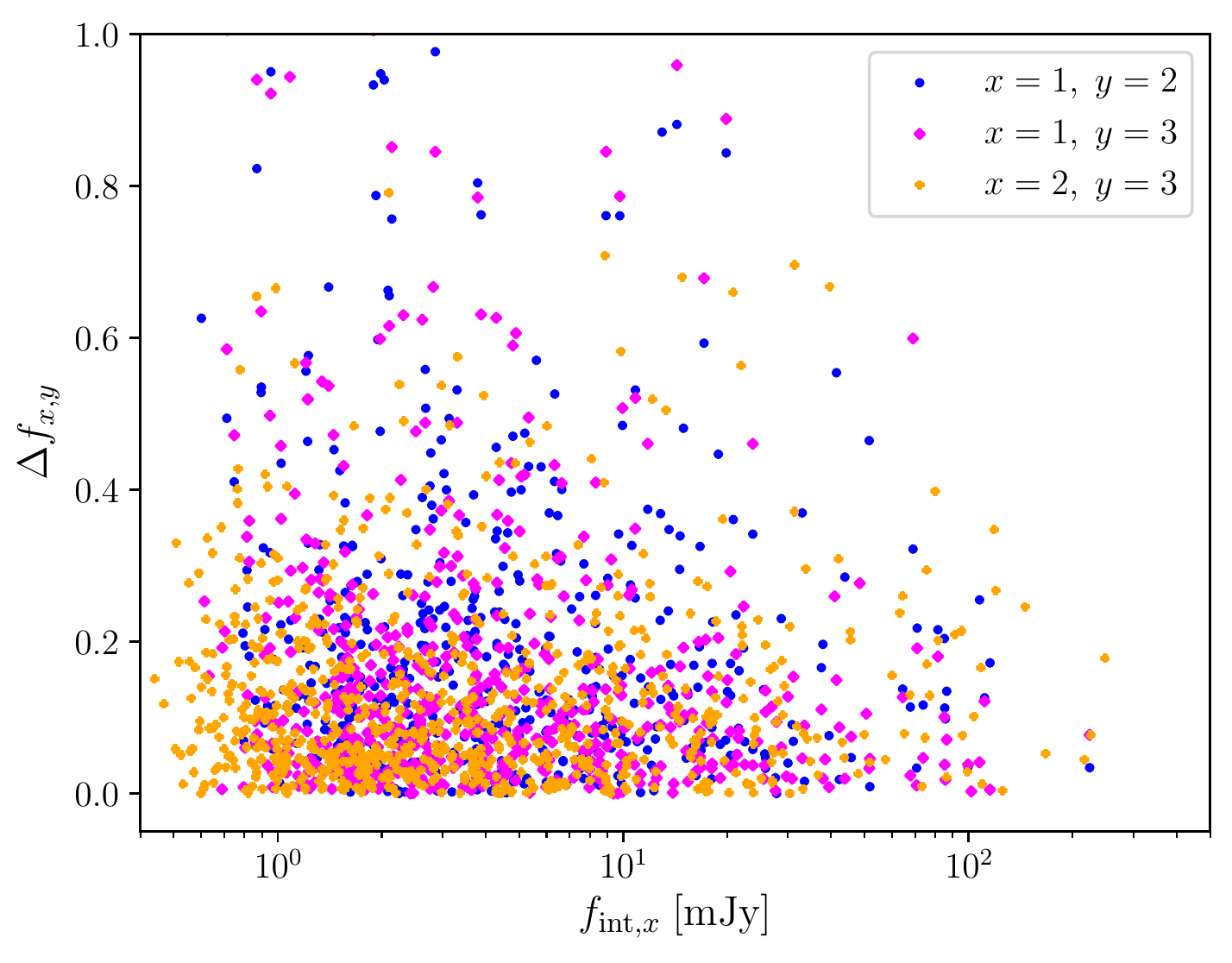}
    \caption{The integrated flux versus the absolute value of the relative difference in flux $\Delta f_{x,y}$, defined in Eq. \ref{eq:reldiff}. This is calculated for epoch 1 and 2 (blue circles), epoch 1 and 3 (pink diamonds), and epoch 2 and 3 (orange crosses). The scatter is larger than what could be expected in terms of rms noise error with median values of $\Delta f_{1,2}$, $\Delta f_{1,3}$ and $\Delta f_{2,3}$ of 13.7\%, 10.6\% and 8.9\%, respectively. We suspect that the main cause of the scatter is the quality of the phase calibration.}
    \label{fig:fluxscat}
\end{figure*}
We observed a target field with Apertif which was chosen to cover a 9.5 $\mathrm{deg}^2$ region of high probability in the localisation skymap.
This field  centred on coordinates $\alpha = 16^\mathrm{h}25^\mathrm{m}12^\mathrm{s}$, $\delta = +17^\circ47{}^\prime24{}^{\prime\prime}$ (J2000) was observed three times for 12 h at $\Delta T =$ 68, 90, 109 d post merger. We will refer to these observations as Epoch 1, 2 and 3, respectively. Table \ref{tab:obssum} gives a summary of the observations. For epoch 2, all twelve Westerbork antennas with Apertif PAFs (on the fixed, equidistant Radio Telescopes RT2..RT9 plus the movable dishes RTA..RTD) were operational. For epoch 1 antenna RTD was not operational, and for epoch 3 antenna RT2 was not in use. Furthermore, data from antenna RTC was not included in the processing of epoch 1 as this dish suffered from delay issues. 

The field is covered by 40 partially overlapping beams formed from the PAFs with a FWHM of$\sim$ 35 arcminutes each on the sky. The data has a centre frequency of 1280 MHz with 300 MHz of bandwidth being recorded. The flux and bandpass calibrator 3C 147 was either observed before (first and third epoch) or after (second epoch) the observation for 3 mins for each of the 40 Apertif beams. The data was imaged for each beam separately using Apercal, 
the Apertif imaging pipeline~\citep{b_adebahr_apercal_2021}, version 2.5.
Due to high levels of Radio Frequency Interference (RFI) in the lower part of the band, only the upper 150 MHz is processed. As the signal is expected to be broad band, the frequency coverage itself is not overly important, but reducing the bandwidth by a factor of two results in an increase of the image noise, and thus the detection threshold, by a factor of $\sqrt{2}$~\citep{e_a_k_adams_first_2021}  

As part of the pipeline, automatic flagging of RFI is performed using AOFlagger~\citep{offringa_post-correlation_2010,offringa_morphological_2012}. This resulted in $\sim$ 14\%, 21\% and 12\% of the processed data per beam being flagged, on average, due to RFI for epoch 1, 2 and 3, respectively. Additionally, the first 12 mins of observation for epoch 2 were flagged manually as the dishes were not properly settled on source. Including this data resulted in artefacts in the continuum image, while flagging it only had a minor impact on the overall sensitivity.  

For epoch 1 and 3, all reduction steps including cross-calibration, direction-independent self calibration, imaging and cleaning, were done using standard pipeline parameters~\citep{b_adebahr_apercal_2021}. In particular, the intervals of the self calibration solutions, used to update the solutions obtained from the calibrator observations, were derived automatically. For epoch 2,  however, solution intervals of 30s, shorter than derived by the pipeline, were necessary to suppress notable phase artefacts in the cleaned continuum images.  

The final data product per beam produced by Apercal is a multi-frequency Stokes I image  created over the full processed frequency range. A mosaicking script then combines the beam continuum images using a linear algorithm. An inverse-squared weighting, based on preliminary primary beam response models of Apertif,  is used to adjust for decreasing sensitivity away from the phase centre. Additionally, different parts of the observation were flagged per beam,  affecting the  resolution. To ensure a consistent resolution across the mosaic, the images are convolved to the common highest possible resolution before making the mosaic. For both epoch 1 and 3, one beam failed internal data quality checks implemented in the pipeline and was not used when creating the mosaics. For epoch 2, nine beams failed internal data quality checks and were not used in the mosaic. As a consequence, epoch 2 has a slightly reduced FOV compared to the other epochs. We show the mosaic of the third epoch in Fig. \ref{fig:mosaic}.

The beam size of epoch 2 is 50 arcsec $\times$ 12 arcsec, with major axis position angle PA$=-0.1^\circ$;  consistent with that at epoch 3, of 48 arcsec $\times$ 11 arcsec (PA$=-0.6^\circ$). The beam size of epoch 1, 90 arcsec $\times$ 23 arcsec (PA$= 0.6^\circ$) , is considerably worse. We there lacked the longest baselines, as data from antennas RTC and RTD could not be used.
The best rms noise sensitivity reached is $\sim$ $70 \ \mu \mathrm{Jy}$, $60 \ \mu \mathrm{Jy}$ and $50 \ \mu \mathrm{Jy}$, for epoch 1, 2 and 3, respectively. This sensitivity decreases closer to the edges of the mosaics or around bright sources due to direction-dependent errors. We inspect the stability of the source fluxes in the next section. 

\section{Flux consistency between epochs}
\label{sec:fluxcons}
The radio sky is relatively quiet compared to other wavelengths, with an estimated transient rate of less than 0.37 $\mathrm{deg}^{-2}$ at
1.4 GHz for sources with a peak flux density greater than 0.21 mJy~\citep{mooley_sensitive_2013}. Furthermore, most sources should not exhibit any variation in their flux besides measurement errors due to noise (see, e.g., ~\citealp{sarbadhicary_chiles_2020} and references therein). Because the observations were taken during the Apertif commissioning phase, it is important to test how stable the measured source fluxes actually are between the epochs. To this extent, the mosaics of each epoch were analysed using the source finder PyBDSF \citep[which is also used in the Apercal pipeline;][]{mohan_pybdsf_2015} and the integrated fluxes for unresolved sources with a positional match within 10 arcsec were compared. We used the local noise map computed by PyBDSF and selected sources 5$\sigma$ above the local rms noise level. Surrounding pixels above the 3$\sigma$ level were also used for source fitting. We constrained the shape of the Gaussian fit to the restoring beam.  

The flux scale is consistent between all epochs, which is within the usual range of flux uncertainties for imaging observations. We recover a median of the ratio in fluxes of 1.00, 0.97 and 0.96 between epoch 1 and 2, epoch 1 and 3, and epoch 2 and 3, respectively. To quantify the scatter in the fluxes, for each source we define:
\begin{equation}
\label{eq:reldiff}
    \Delta f_{x,y} = \Big|\frac{f_{\mathrm{int},x}-f_{\mathrm{int},y}}{f_{\mathrm{int},x}}\Big|,
\end{equation}
where $\Delta f_{x,y}$ is the absolute value of the relative difference in integrated flux and $f_{\mathrm{int},x}$ and $f_{\mathrm{int},y}$ are the integrated fluxes in epoch $x$ and $y$.

The scatter in the radio fluxes is higher than expected, with the median of $\Delta f_{1,2}$, $\Delta f_{1,3}$ and $\Delta f_{2,3}$ being 13.7\%, 10.6\% and 8.9\%, respectively. We plot this statistic as a function of integrated flux in Fig. \ref{fig:fluxscat} for each combination of epochs. The scatter does not exhibit a strong flux dependence beyond what could be expected in terms of rms noise errors. We suspect that the quality of the phase calibration is the main source of this scatter. We discuss these errors, other possible causes, and the implications in our search for the afterglow in the next section.
\section{Search for the afterglow of GW190425}\label{sec:transientsearch}
\subsection{Radio variables and transients search}
We made use of the LOFAR Transients Pipeline \citep[TraP;][]{swinbank_lofar_2015} to detect variable sources or transients in our observations. TraP identifies sources over multiple epochs and constructs light curves for each source. From these light curves, statistics are calculated which can be used to quantify the variability of sources. For sources which are only identified in one epoch, a light curve cannot be constructed. Instead, using TraP we assigned these sources a candidate transient classification based on their signal-to-noise-ratio (SNR) properties. For a detailed explanation of TraP and the associated variability statistics we refer the reader to~\citealp{swinbank_lofar_2015}.

We set up TraP such that pixels with a flux count 5$\sigma$ above the local rms noise level are identified as a source and surrounding pixels with a flux count 3$\sigma$ above the noise become a part of that same source. The Gaussian fit that is then performed on the source to measure its flux density is constrained to have the same shape as the restoring beam. The dimensionless \textit{de Ruiter} radius~\citep{de_ruiter_westerbork_1977}, which is used to associate sources between epochs based on their angular separation, is set to the default value of $5.68$. Furthermore, matching sources between epochs should have an angular separation of less than one semi-major axis of the restoring beam.

For sources with flux density measurements in two or three epochs, the reduced chi-squared statistic $\eta$ and the fractional variability statistic $V$ are calculated by TraP from the resulting light curve. A Gaussian function is fit to the distribution of both statistics in logarithmic space with mean and standard deviation $\mu_{V}$, $\sigma_{V}$ and $\mu_{\eta}$, $\sigma_{\eta}$, respectively.  Sources with both $\eta$ and $V$ significantly higher than the mean values of their distributions are identified as candidate variable sources. Thresholds for both metrics to separate such variables from stable sources can be chosen somewhat freely. Different thresholds do influence recall and precision rates of the produced set of candidate variables considerably~\citep{rowlinson_identifying_2019}. We followed~\citet{dobie_askap_2019} and specified a source as candidate variable if both  $\eta > \mu_{\eta} + 1.5\sigma_{\eta}$ and $V > \mu_{V} + 1\sigma_{V}$.

\subsection{Expected afterglow emission of GW190425}
\begin{table*}
\begin{tabular}{@{}lllllllll@{}}
\hhline{=========}
R.A. (deg) & Decl. (deg) & $f_{\mathrm{int},1}$ (mJy) & $f_{\mathrm{int},2}$ (mJy) & $f_{\mathrm{int},3}$ (mJy) & $f_{\mathrm{int},2}/f_{\mathrm{int},1}$ & $f_{\mathrm{int},3}/f_{\mathrm{int},2}$ & $V$  & $\eta$  \\ \hline
246.000    & 18.824      & 1.93 $\pm$ 0.136        & 3.68 $\pm$ 0.134        & 3.82 $\pm$  0.104   & 1.91 $\pm$ 0.152   & 1.04 $\pm$ 0.0472   & 0.34 & 67.55   \\
247.214    & 18.471      & 1.95 $\pm$ 0.170        & 3.15  $\pm$ 0.146       & 4.10 $\pm$ 0.102    & 1.62 $\pm$ 0.160 & 1.29 $\pm$ 0.0679  & 0.35 & 60.07   \\
247.354    & 18.527      & 12.9 $\pm$ 0.179       & 24.3 $\pm$ 0.153       & 24.3 $\pm$ 0.113   & 1.89 $\pm$ 0.0286 & 1.00 $\pm$ 0.00780 & 0.32 & 1607.35 \\ 
247.330    & 18.682      & 14.3 $\pm$ 0.155       & 27.0 $\pm$ 0.190       & 26.8 $\pm$ 0.136   & 1.89 $\pm$ 0.0244 & 0.99 $\pm$ 0.00857 & 0.32 & 2183.47 \\
247.233    & 18.316      & 20.2 $\pm$ 0.193       & 33.3 $\pm$ 0.190       & 32.6 $\pm$ 0.124   & 1.65 $\pm$ 0.0184 & 0.98 $\pm$ 0.00671 & 0.26 & 1668.20 \\ \hline
\end{tabular}
\caption{Overview of the five candidate variables found in our observations with flux measurements consistent with the expected afterglow emission of GW190425. After cross matching these candidates with the GLADE catalogue ~\citep{dalya_glade_2018}, no possible host galaxies were found within 3$\sigma$ of the estimated luminosity distance of GW190425. None of these candidates are thus a source of the afterglow emission. We suspect that for these candidates their variability is either caused by direction-dependent errors from a nearby bright source (row two to five), or insufficient phase calibration (first row).}
\label{tab:varcands}
\end{table*}
To search for the radio afterglow of GW190425, we aim to select the potential candidates identified by TraP that have a flux evolution consistent with the expected afterglow emission. We based the criteria for emission partly on the light curve of the sole radio counterpart detected to date: that of GW170817. 
That event peaked at $\Delta T = 174^{+9}_{-6} \ \mathrm{d}$~\citep{mooley_strong_2018}, significantly later than our  last epoch at $\Delta T = 109 \ \mathrm{d}$ post merger.
In population studies~\citep{duque_radio_2019}, however, a significant fraction of BNS-merger afterglows peak closer to our first or second epoch. 
Given this range, peak time is not a criterion for candidate selection per se. 
Even so, these jet dominated afterglows should display only a single rise and decline in flux~\citep{hotokezaka_radio_2016,dobie_askap_2019},
as observed in GW170817~\citep{mooley_strong_2018,mooley_superluminal_2018,ghirlanda_compact_2019}. 
We therefore excluded sources with both a decrease in flux density from epoch 1 to 2 and an increase in flux density from epoch 2 to 3.

If the afterglow emission of GW190425 exhibits the same flux evolution as GW170817, the associated radio source should be selected by TraP as an outlier in flux variability compared to the general population of radio sources. The scatter in fluxes mentioned in Section \ref{sec:fluxcons}, however, might hinder our ability to detect the afterglow emission. We can estimate the expected increase in flux density using the light curve of GW170817~\citep{mooley_strong_2018}.
The radio flux would have increased by $\sim$ 26 \% between the time of our epoch 1 and 2 and by $\sim$ 17\% between the time of our epoch 2 and 3. These increases are still larger than the median scatter in our observations. Some models predict even stronger changes in flux over the timescale of our observations~\citep{hotokezaka_radio_2016}. Thus, we feel confident that a radio source of the afterglow of GW190425 should still be picked up as an outlier in variability. Even so, the significance of such a detection will be low. We show the results of our search next.
\subsection{Candidate variables}
\label{sec:canvar}
We obtained the following fit to the distributions of $V$ and $\eta$ in logarithmic space: $\mu_{V} = -0.97$, $\sigma_{V} = 0.38$ and $\mu_{\eta} = 0.16$, $\sigma_{\eta} = 1.07$. This corresponds to thresholds of $\eta>59$ and $V>0.25$. Using these thresholds, we recovered 30 candidate variables in our observations. From the 30 candidate variables, we selected those sources with flux measurements consistent with the expected afterglow emission. This resulted in 25 candidate variables possibly associated with GW190425 in our observations. After further manual inspection, 20 candidates in regions of increased noise, e.g., close to the edges of the mosaics, were found to have unreliable flux measurements and were discarded.

The five remaining candidates were cross matched with the GLADE catalogue~\citep{dalya_glade_2018} within a 30 arcsec radius (corresponding to $\lesssim 20$ kpc offset) to look for possible host galaxies. No galaxies were found within 3$\sigma$ of the estimated luminosity distance $159^{+69}_{-72}$ Mpc (90\% credible intervals) of GW190425~\citep{abbott_gw190425_2020}. We conclude that none of these five candidates are a source of the afterglow emission of GW190425. Possible other origins of their variability are discussed in Section \ref{sec:trapdisc}. For completeness, we list their properties in Table \ref{tab:varcands}.

\subsection{Candidate transients}
Sources with only one flux density measurement can appear in either one of the three epochs to be consistent with the expected afterglow emission. TraP found 291 such sources in our observations, all in the third epoch. Upon manual inspection, most of these sources do seem to be present in previous epochs. We suspect that they were not identified as a source due to their shape not matching the restoring beam. 
We checked this hypothesis by running TraP without the previously mentioned restoring beam constraint. This led to substantially more sources being properly recognised in the first two epochs and not identified as a new source in the third epoch. However, many more image artefacts were now incorrectly being identified as a new source instead. We thus opted to continue with the restoring beam constraint in place. 

To filter out most sources which were either in an area with high noise or had a shape inconsistent with the restoring beam in previous epochs, we made use of their SNR calculated by TraP in the best noise region of the epoch with the previous lowest rms value~\citep{swinbank_lofar_2015}. If the SNR in the best noise region crossed our detection threshold plus some extra margin $M$, the source was determined to be a candidate transient. What margin $M$ to choose depends both on the quality of the observations and desired recall and precision rates of the set of possible transients (see~\citealp{rowlinson_identifying_2019} for an in depth discussion). Setting $M = 34$~\citep{rowlinson_identifying_2019} worked well to cut out most easily identifiable false positives in our candidate transient set. This narrowed down our search to 55 transient candidates. After manual inspection, all candidates were determined to result from either direction-dependent errors around bright sources or noise artefacts at the edges of the mosaics. Thus, we identified no transients in our observations as a source of the afterglow emission of GW190425.    

\subsection{Discussion}
\label{sec:trapdisc}
\begin{figure*}[t]
    \centering
    \mbox{\includegraphics[width=0.48\textwidth]{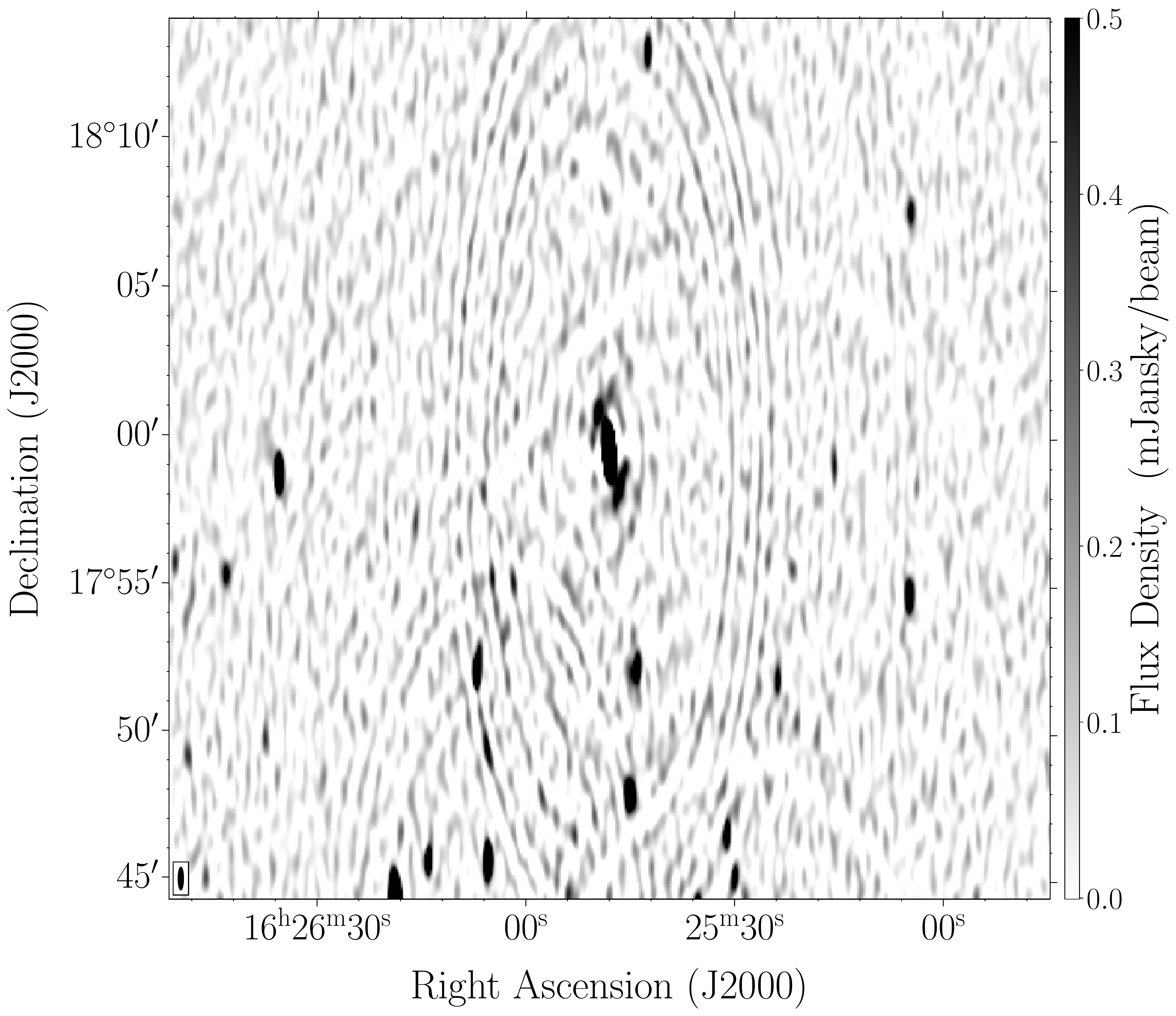}}
    \mbox{\includegraphics[width=0.48\textwidth]{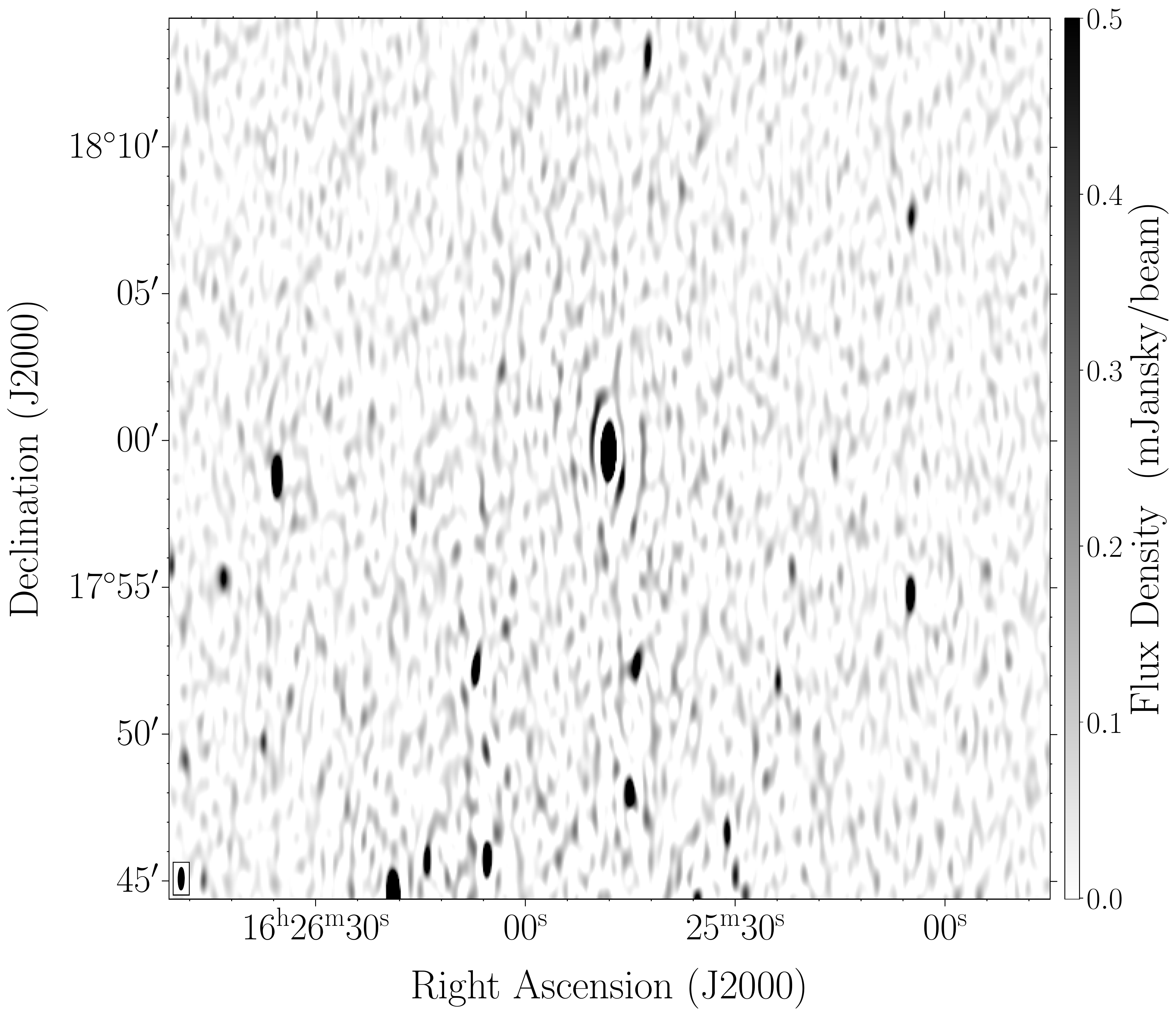}}
    \caption{Continuum image for beam "00" of epoch 2 made with Apercal (left) and manually reprocessed (right, see text for details). Image quality is improved in the reprocessed image with better sidelobe suppression of the central bright source and less visible rings from direction-dependent errors.}
    \label{fig:beam00}
\end{figure*}
The ideal observing campaign for detecting a radio counterpart to GW190425 in our field of view 
involves a considerable number of repeat visits, and an image quality approaching the theoretical sensitivity limit. 
The study we present here did not yet meet these specifications. 
Below will describe the factors that limited our ability to detect the afterglow.  

For sources with flux density measurements in only a few epochs, the reduced chi-squared statistic $\eta$ can vary appreciably over these epochs~\citep{rowlinson_identifying_2019}. In general, the three epochs we observed might not be enough for most sources to robustly determine this measure for the stability of their light curve. Furthermore, this stability is also impacted by the general systematic variability mentioned in Section \ref{sec:fluxcons}. 
Outliers in variability are less apparent when the scatter in flux between observations is already substantial for most sources.
Thus, the distribution of $\eta$ in log-space becomes wider, which is apparent in the large value of the standard deviation $\sigma_{\eta}$ from the Gaussian fit to $\eta$. For reference, Fig \ref{fig:varstat} in appendix \ref{sec:appstat} shows a histogram of the 2-dimensional distribution of $\eta$ and $V$ with the large spread in $\eta$ clearly visible.

We presume the source of the general systematic variability in our observations to be related to the data reduction. We deem the main source of error to be the quality of the phase calibration. If the phases are not sufficiently calibrated, the emission from point sources is not actually point-like but partially spread out. This can affect the flux that is measured in the source-finding procedures. Besides the resulting scatter in fluxes between the epochs, these calibration errors were also apparent when using TraP to find transients. Many sources were not properly recognised across the epochs due to their inconsistent shape. This produced many false positives in our candidate transient set. 

Uncertainties in the observations for the primary beam models could also increase the scatter in the flux measurements. Two possible complications are, for example, strong RFI during these observations and the malfunctioning of individual elements. Strong RFI could interfere with the measurements of the primary beam response. The frequency range used in our primary beam model, however, is normally relatively RFI free. We thus do not expect this to be a big influence. Some antennas might have a deformed primary beam shape because of malfunctioning individual elements. This may not be taken into account in the global primary beam model used which is averaged across all antennas. The above, and other, factors can change on timescales shorter than the time between our epochs. 

The influence of such variations in the mosaics would be partially compensated at low levels of the primary beam response. Here, the beams overlap which suppresses model errors. Strong changes in the inner part of the beam would not be compensated as much. For epoch 1 and 2, the same set of primary beam models, closest in observation time to these epochs, were used. Epoch 3 was corrected with a different set of models. To measure any fluctuations between the two sets of models, we made a mosaic of the primary beam weights for each set. Between the two mosaics, the difference is at most $\sim$ 4\% at the centre of the beams, with variations often below 1\% at the edges. These fluctuations are thus too small to be the main source of the systematic variability in flux.

Sources which are near very bright sources are additionally affected by their direction-dependent errors. These errors gave rise to a non-Gaussian distribution of the noise in parts of our images which impeded accurate flux measurements. We suspect that for four of the five candidates in Section \ref{sec:canvar} their variability is due to these errors. They are relatively close to the same bright source and their relative change in flux between each epoch is similar, see row two to five in Table \ref{tab:varcands}. The other candidate, first row in Table \ref{tab:varcands}, also has a similar light curve evolution but is not particularly close to the other candidates. The shape of this source deviates from the restoring beam in epoch 1, likely due to the insufficient phase calibration,
and presumably does not have its flux accurately measured in this epoch. As the flux is constant within the errors between epoch 2 and 3, we believe that the variability of this source, evident in the large values for $\eta$ and the fractional variability $V$, is a result of the calibration error in epoch 1. The characteristic rings resulting from the direction dependent errors also triggered a lot of false positives in the candidate transients set. Reducing these errors is an active area of development for Apercal. In Section \ref{sec:impimg}, we detail a manual reprocessing of a few Apertif beams to improve the image quality.

In summary, a number of aspects of our search can be improved for future follow-up campaigns.
Their impact would foremost be to reduce the number of false-positive candidates.
Any sufficiently bright transients in our observing field would still have been detected.

\section{Future observing prospects}
\label{sec:obsprosp}
\subsection{Improvements in Apertif Imaging}
\label{sec:impimg}
The success of future afterglow searches with Apertif is partly contingent on the achievable image quality. The first Apertif survey data release~\citep{e_a_k_adams_first_2021} has shown that noise levels down to 30 $\mu \mathrm{Jy}$ and good image quality are regularly attained using automatic Apercal pipeline processing. 
While the automatic processing did not reach a similar quality for our observations, future reprocessing of the data could potentially yield improvements with updated versions of Apercal\footnote{Work is ongoing to improve the self calibration of the phases in Apercal and implement direction-dependent calibration.}.

To investigate this prospect, we reprocessed a few centre Apertif beams as a demonstration, using WSClean~\citep{offringa_wsclean_2014} and DPPP~\citep{van_diepen_dppp_2018}. First, a further cross-calibration was applied to the data using a mask derived from the NVSS survey~\citep{condon_nrao_1998}. This was followed by two cycles of direction-independent calibration and imaging in a standard way on minute solution intervals. Then the sources with strong artefacts were identified in the images, and the corresponding directions were stored. Finally, additional direction-dependent calibration was performed in these directions to solve for both amplitude and phase variations on an hour solution interval.

We show a cut-out of the continuum image for beam "00" of epoch 2 in Fig. \ref{fig:beam00}. The image on the left was made with Apercal, while the image on the right was manually reprocessed as described. Compared to Apercal, the sidelobes of the bright source in the centre are better suppressed in the reprocessed image. Furthermore, the rings from the direction-dependent errors around this source are nearly gone. Noise levels are improved reaching $\sim$ 45 - 50 $\mu \mathrm{Jy}$ in the centre of the reprocessed image. For epoch 3, the manual reprocessing of beam "00" gives similar improvements, while for epoch 1 the difference with Apercal is less pronounced. 

For bright sources with integrated fluxes above a few mJy, the fluxes were consistent within 5\% in the reprocessed images of epoch 2 and 3 for beam "00". Compared to the $\sim$ 10\% scatter in the Apercal images for such sources, this is a noticeable improvement which points to the source shapes being more consistent between the two epochs. It is difficult, however, to measure the overall scatter in flux based on a single beam as the number of faint sources is limited. To measure if the scatter decreased for faint sources too, we did an initial comparison in flux between mosaics of seven reprocessed centre beams for epochs 2 and 3. No major changes in the scatter were present in the integrated flux between the epochs compared to the results of Section \ref{sec:fluxcons}. Of the few reprocessed beams, beam "00" had the most obvious improvements in image quality. We thus did not find a universal improvement in image quality across beams. We reason improvements are unlikely when running the reprocessed data through TraP to search for variable sources. Still, the direction-dependent calibration does reduce artefacts. 
This will lead to fewer false positive transients in our future afterglow searches.

We conclude that more observational epochs, further characterisation and understanding of the Apertif system, and improvements in the pipeline will greatly benefit our search for radio counterparts to BNS mergers. In the next section, we will describe the prospects for finding these counterparts in the fourth and future observing runs of the gravitational-wave detector network.

\subsection{Expanding GW network of detectors}
\label{sec:expnet}
The fourth observing run (O4) of the gravitational-wave detector network is expected to commence in 2022, although the early suspension of O3 has made the start date and the anticipated detector sensitivities more uncertain. Here, we will use the numbers for sensitivities outlined by the KAGRA Collaboration, LIGO Scientific Collaboration and Virgo Collaboration~\citep{abbott_prospects_2020}. Four detectors are planned to be operational for 1 year in O4: the two Advanced LIGO (aLIGO) detectors at design sensitivity, Advanced Virgo (AdV) in its Phase 1 upgrade and the newest addition, KAGRA~\citep{somiya_detector_2012,the_kagra_collaboration_interferometer_2013}. During this year, the aLIGO detectors, AdV and KAGRA will have an anticipated BNS range $R$ of 160-190 Mpc, 90-120 Mpc and 25-130 Mpc, respectively. Importantly, from the $10^{+52}_{-10}$ BNS events that are expected in total, 38-44\% of these events are predicted to have a 90\% credible region on the sky smaller than 20 $\mathrm{deg}^2$. This can be covered with just three Apertif pointings. While the localisation region on sky will likely be drastically reduced, the increase in BNS range will mean that more events will be detected at larger luminosity distances in O4. This will decrease our ability to detect the radio afterglow due to the inverse square relation between the radio flux and the luminosity distance.  

\begin{figure}
    \centering
    \includegraphics[width=0.48\textwidth]{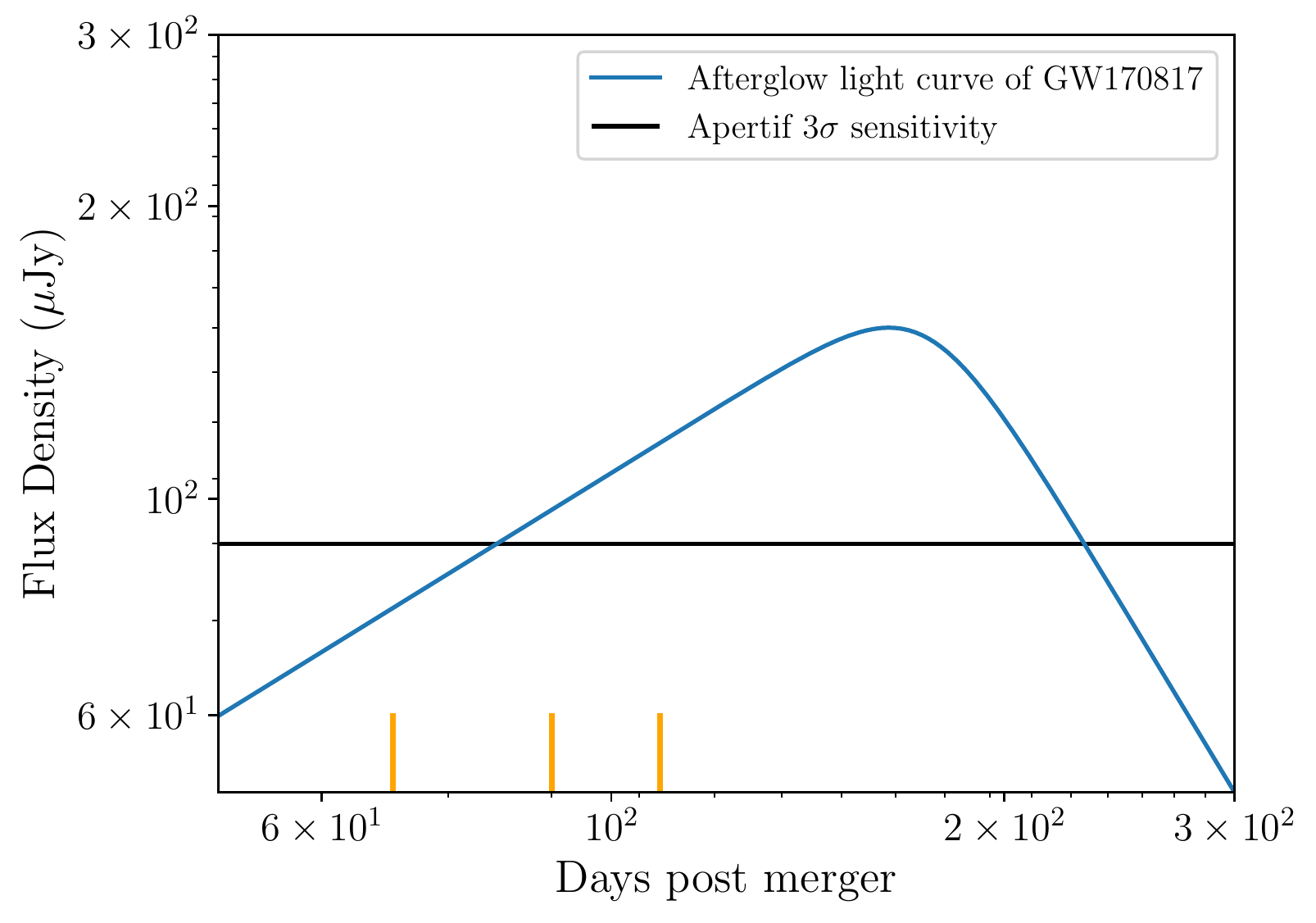}
    \caption{A fit to the radio light curve of the afterglow of GW170817~\citep{mooley_strong_2018} in blue. The curve is plotted on a log-log scale to easily show the power-law dependence. Shown in black is the 3$\sigma$ Apertif design sensitivity. A detection with Apertif would have been possible if it was operational. If we had taken three observations at the same days post merger as our observations for GW190425, orange stripes, we would have detected the afterglow in the last two observations.}
    \label{fig:lcgw170817}
\end{figure}

As a point of reference, we show a fit to the radio light curve of the afterglow of GW170817~\citep{mooley_strong_2018} in Fig. \ref{fig:lcgw170817} in blue with the best achieved 3$\sigma$ Apertif sensitivity indicated in black. We would have been able to detect the emission if Apertif was operational at the time but this would not have been possible if the merger happened much further out. Just one observation would still yield a rather marginal detection. It is thus essential to regularly sample the light curve across multiple epochs to obtain a robust detection with physical information. Especially the peak and the subsequent decay of the light curve should be monitored carefully. Certainly, the decay rate gives crucial information about the physical processes of the BNS merger (see~\citealp{mooley_strong_2018} and references therein). If we had observed GW170817 at the same days post merger as our observations for GW190425, orange stripes in Fig. \ref{fig:lcgw170817}, we would have detected the afterglow in the last two observations.    

To quantify our ability to detect radio counterparts to BNS events in O4 and beyond, we follow the population study in~\citet{duque_radio_2019} and include Apertif. We will summarise both the criterion for radio detection and gravitational-wave detection of the merger in the next section. We refer to their work for a detailed explanation of the afterglow model and the distribution of BNS population parameters.
\subsection{Forecasts for radio detections of binary neutron star merger afterglows}
\begin{figure*}[ht]
    \centering
    \includegraphics[width=0.78\textwidth]{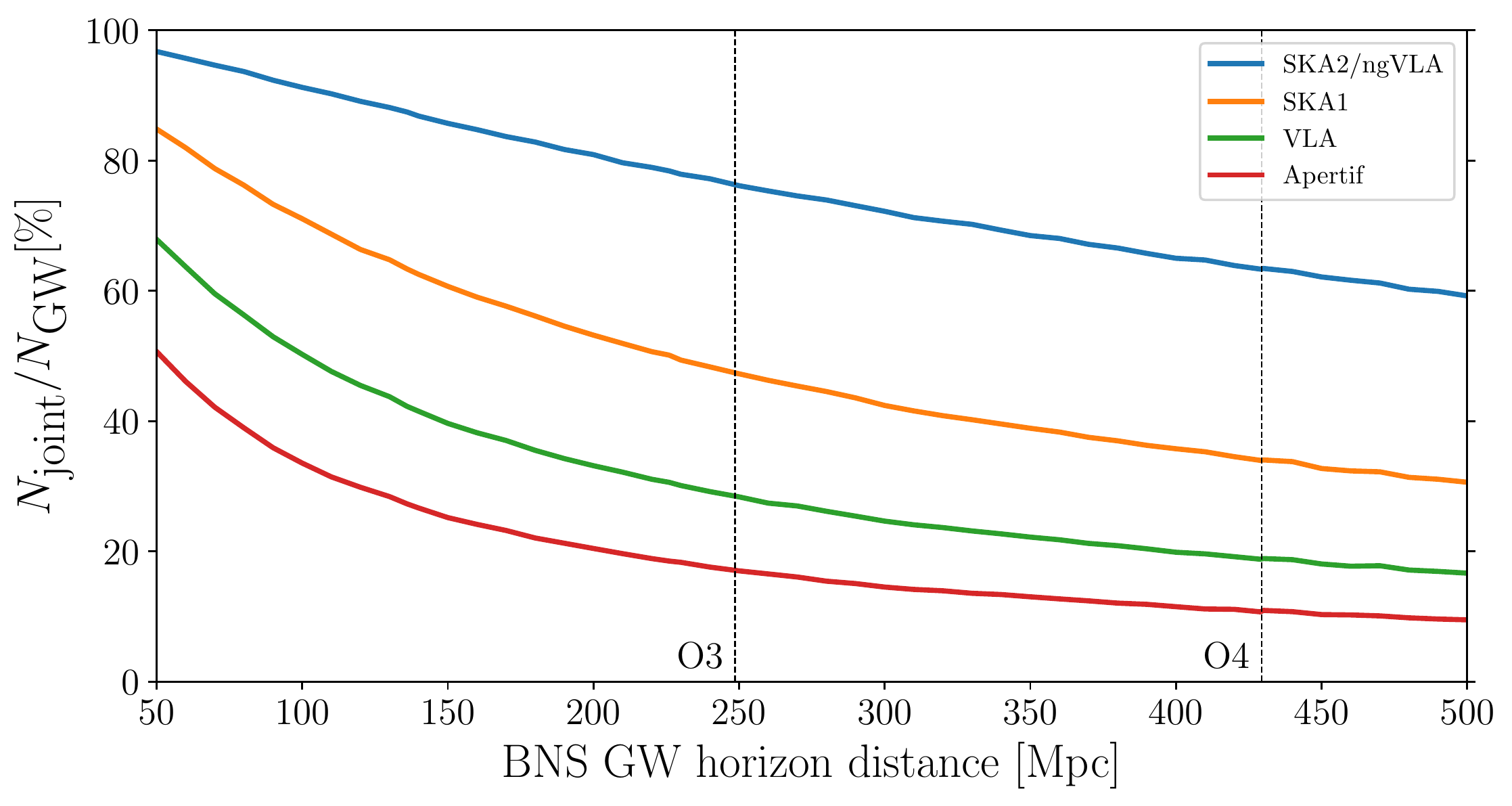}
    \caption{Fraction of BNS mergers that will be detectable both in GWs and through their radio afterglow ($N_{\mathrm{joint}}$) versus those only detectable in GWs ($N_{\mathrm{GW}}$). The mergers are simulated up to the horizon of aLIGO Hanford with the dashed vertical lines indicating the sensitivities in O3 and O4. SKA2 or ngVLA, SKA1, VLA and Apertif are shown as the blue, orange, green and red curves, respectively. During O4, Apertif will be able to detect $\sim$ 11 \% of the well localised GW events.}
    \label{fig:popstudy}
\end{figure*}
It is assumed that the peak of the BNS afterglow emission is dominated by a relativistic core jet with the peak flux scaling as~\citep{nakar_detectability_2002}:
\begin{equation}
F_{\mathrm{p},\nu} \propto E_{\mathrm{iso,c}} \ \theta_{j}^{2} \ n^{\frac{p+1}{4}} \ \epsilon_{\mathrm{e}}^{p-1} \ \epsilon_{\mathrm{B}}^{\frac{p+1}{4}} \ \nu^{\frac{1-p}{2}} \ D^{-2} \ \mathrm{max}\Big(\theta_j,\theta_v \Big)^{-2p},    
\end{equation}
where $E_{\mathrm{iso,c}}$ is the core jet isotropic equivalent kinetic energy, $\theta_{j}$ is the jet opening angle, $n$ is the external density, $\epsilon_{\mathrm{e}}$, $\epsilon_{\mathrm{B}}$ and $p$ are shock microphysics parameters, $\nu$ is the wavelength, $D$ is the distance and $\theta_v$ is the viewing angle. The afterglow was assumed to be detectable if the peak flux exceeded the sensitivity of the radio telescope $s$ which we set at $90  \ \mu \mathrm{Jy}$ for Apertif observing at $\nu = 1.4 \ \mathrm{Ghz}$. As a comparison, we also included four radio arrays at sensitivities listed in~\citet{duque_radio_2019} observing at $\nu = 3 \ \mathrm{Ghz}$: the Karl Jansky Very Large Array (VLA) at $s = 15 \ \mu \mathrm{Jy}$, the Square Kilometer Array~\citep{fender_transient_2015} in phase 1 (SKA1) at $s = 3 \ \mu \mathrm{Jy}$, and SKA in phase 2 (SKA2) and Next Generation VLA (ngVLA, \citealp{selina_next-generation_2018}) both at $s = 0.3 \ \mu \mathrm{Jy}$.

The BNS mergers detected and localised in GWs were determined using the following single detector criterion:
\begin{equation}
\label{eq:crit}
    \sqrt{\frac{1+6 \cos^2 \theta_v + \cos^4 \theta_v}{8}} > \frac{D}{\Bar{H}},
\end{equation}
where $\bar{H} = \sqrt{\frac{2}{5}}H$, with $H = 2.26 R$ the horizon of the second most sensitive instrument in the detector network, in this case aLIGO Hanford. Using this criterion, it was presumed that if aLIGO Hanford detected the GW, it was also detected by aLIGO Livingston (the most sensitive instrument in the network). We also assumed that this led to a sufficiently small localisation region for radio follow-up to be possible with one or a few telescope pointings. We note that this is a relatively simple criterion which might underestimate the size of the localisation region inferred solely through GW detection. 

For the Monte Carlo simulation we used the fiducial population model from~\citet[][their Section 4.1]{duque_radio_2019}. 
This uses the following parameter distributions: 
a broken power-law distribution  for $E_{\mathrm{iso,c}}$, motivated by the gamma-ray luminosity function of cosmological short gamma-ray bursts~\citep{beniamini_revised_2016,duque_radio_2019}, with a density of probability:
\begin{equation}
\phi(E_{\mathrm{iso,c}}) \propto
\begin{cases}
&E_{\mathrm{iso,c}}^{-\alpha_1} \  \mathrm{for} \ E_{\mathrm{min}} \leq E_{\mathrm{iso,c}} \leq E_{\mathrm{b}} \\
&E_{\mathrm{iso,c}}^{-\alpha_2} \  \mathrm{for} \ E_{\mathrm{b}} \leq E_{\mathrm{iso,c}} \leq E_{\mathrm{max}} 
\end{cases}
\end{equation}
where $\phi$ is normalised to unity, $E_{\mathrm{min}} = 10^{50} \ \mathrm{erg}$ and $E_{\mathrm{max}} = 10^{53} \ \mathrm{erg}$. Furthermore,  $\alpha_1 = 0.53 $, $\alpha_2 = 3.4 $ and $E_{\mathrm{b}} = 2 \times 10^{52} \ \mathrm{erg}$ were used,  adopted from~\citet{ghirlanda_short_2016}. For both $n$ and $\epsilon_{\mathrm{B}}$ a log-normal distribution was used with central value $\mu = 10^{-3}$ and standard deviation $\sigma = 0.75$. Similar values have been reported when fitting the afterglow of GW170817 (see, e.g.,~\citet{hallinan_radio_2017}). The distribution was restricted to $[10^{-4}, 10^{-2}]$ for $\epsilon_{\mathrm{B}}$.   The rest of the jet parameters were fixed at typical values of $\theta_{j} = 0.1 \ \mathrm{rad}$, $\epsilon_{\mathrm{e}} = 0.1$, and $p = 2.2$. The binaries were distributed uniformly in volume and $\cos \theta_v$.

We simulated a sufficient amount of binaries within the sky-averaged horizon distance $\Bar{H}$ for convergence. In Fig. \ref{fig:popstudy}, we plot the ratio of events detected both in GWs and in radio through their afterglow emission ($N_{\mathrm{joint}}$) versus those only detected in GWs ($N_{\mathrm{GW}}$). We repeated the simulation multiple times to cover a range in $H$ beyond the expected detector sensitivity of O4. 
\subsection{Monte Carlo simulation}
Our results from the Monte Carlo simulation for the observatories besides Apertif are nearly identical to~\citet{duque_radio_2019}, confirming our methods. Future observatories such as SKA1, SKA2 or ngVLA, which are not yet operational during O4, could be sensitive enough to facilitate population level studies of BNS afterglows. They pose great promise in future observing runs. If they were operational during O4 already, SKA1 would detect 34\% of the well localised events with SKA2 detecting almost two-thirds of the events at 64\% coverage. 

From the telescopes that will be operational during O4, Apertif will be able to detect roughly a third of the afterglows that SKA1 would detect. This is $\sim$ 11\% of the well localised GW events. The VLA will be able to detect $\sim$ 19\% of the same events. While Apertif is certainly less sensitive than the VLA, the criterion of Eq. \ref{eq:crit} does not take into account the field-of-view of the different arrays. In this regard, Apertif has an advantage over the VLA. Hence, for events with bigger localisation regions it could be that Eq. \ref{eq:crit} is not met, impeding a VLA observation of the afterglow, whereas Apertif might still be able to make a detection.
Furthermore, even if Eq. \ref{eq:crit} is met, the localisation region may still only be tractable for follow-up by large field-of-view telescopes. Previous GW detections by only two detectors have shown large localisation regions in the past (Fig. 5 of \citealp{abbott_prospects_2020}) but we note that optical/infrared follow up could also be crucial for localising such events.  

Half of the BNS events in O4 will likely have a localisation small enough to be followed up with Apertif. The median 90\% credible localisation region is just 33 $\mathrm{deg}^2$~\citep{abbott_prospects_2020}, or about four Apertif pointings if the shape matches. From the expected BNS events in O4 and the results presented above, it follows that up to three afterglows will be detectable by Apertif. In the optimistic scenario that all events will have either a small localisation region or are localised through, e.g, optical/infrared follow up, this number doubles. Moreover, for the afterglows which have not been observed in optical or infrared wavelengths, for example, for the large fraction that happen in the day-time sky, wide field radio telescopes such as Apertif may be the only way to detect the electromagnetic signal at all. 

As a final point, we emphasise the need for regular radio follow up of the BNS merger. The radio criterion used in these simulations only looks at the detectability of the afterglow. This does not necessarily translate into an actual detection if the emission flux density is close to the sensitivity limit of the telescope (this is also discussed in~\citealp{duque_radio_2019}). The forecasts given in this section should therefore be seen as an upper limit. Even so, multiple marginal detections would still have significant scientific potential and should be actively pursued.   

\section{Conclusion}\label{sec:conclusion}
In this work, we have described the first follow-up of a BNS merger with the new Apertif PAFs installed on the WSRT. We covered an 9.5 $\mathrm{deg}^2$ region of high probability in the localisation skymap of the first O3 BNS event, GW190425, over three observational epochs. While we only observed a small fraction of the approximate 7500 $\mathrm{deg}^2$ 90\% credible localisation region, a possible counterpart could have been of high scientific significance. We identified five candidate counterparts based on their flux variability in our observations. As we found no associated host galaxies for either of the sources at a luminosity distance consistent with GW190425, we ruled them all out. We also looked for possible counterparts in transient sources with only one flux measurement. Our initial analysis found 55 such candidate transients. These were all determined to be imaging artefacts after manual inspection. Although we did not find a radio afterglow counterpart to GW190425, this search helped develop the pipeline and methods which will be instrumental for future searches of radio afterglows with Apertif. Furthermore, the average sensitivity that will be achieved in future observations for finding BNS afterglows will increase with further characterisation and commissioning of Apertif.

We also made predictions for our ability to detect BNS merger afterglows in the fourth observing run of the GW detector network. We extended  simulations from \citet{duque_radio_2019} by  including Apertif and estimate that up to three afterglows will be detectable by the telescope. While the sensitivity of Apertif is lower than other radio telescopes, it has a significant advantage in terms of field-of-view. We caution that considerable uncertainties remain in both the GW localisation array as well as the BNS population distribution. 

\begin{acknowledgements}
We thank Antonia Rowlinson, Mark Kuiack and Kelly Gourdji  
for the helpful discussions and suggestions about TraP. We thank David Gardenier for a helpful discussion about the Monte Carlo simulations. We thank the anonymous referee for their thoughtful comments which have certainly improved this work. This research was supported by Vici research program 'ARGO' with project number 639.043.815, financed by the Dutch Research Council (NWO).
JVL, YM, LCO and RS furthermore acknowledge funding from the European Research Council under the 
European Union's Seventh Framework Programme (FP/2007-2013)/ERC Grant Agreement No. 617199 (`ALERT'), while JMvdH acknowledges ERC funding from (FP/2007-2013)/ERC Grant Agreement No. 291531 (‘HIStoryNU’).
EAKA is supported by the WISE research programme, which is financed by NWO.
DV acknowledges support from the Netherlands eScience Center (NLeSC) under grant ASDI.15.406.
We make use of data from the Apertif system installed at the Westerbork Synthesis Radio Telescope owned by ASTRON. 
ASTRON, the Netherlands Institute for Radio Astronomy, is an institute of NWO.

\end{acknowledgements}
\section*{Data Availability}
Data used to plot the images in this work is uploaded at:

\url{http://doi.org/10.5281/zenodo.4672444}
\bibliographystyle{yahapj}
\bibliography{ref_final}
\appendix
\section{Distribution of variability statistics}
\label{sec:appstat}
\begin{figure}[ht]
    \centering
    \includegraphics[width=0.48\textwidth]{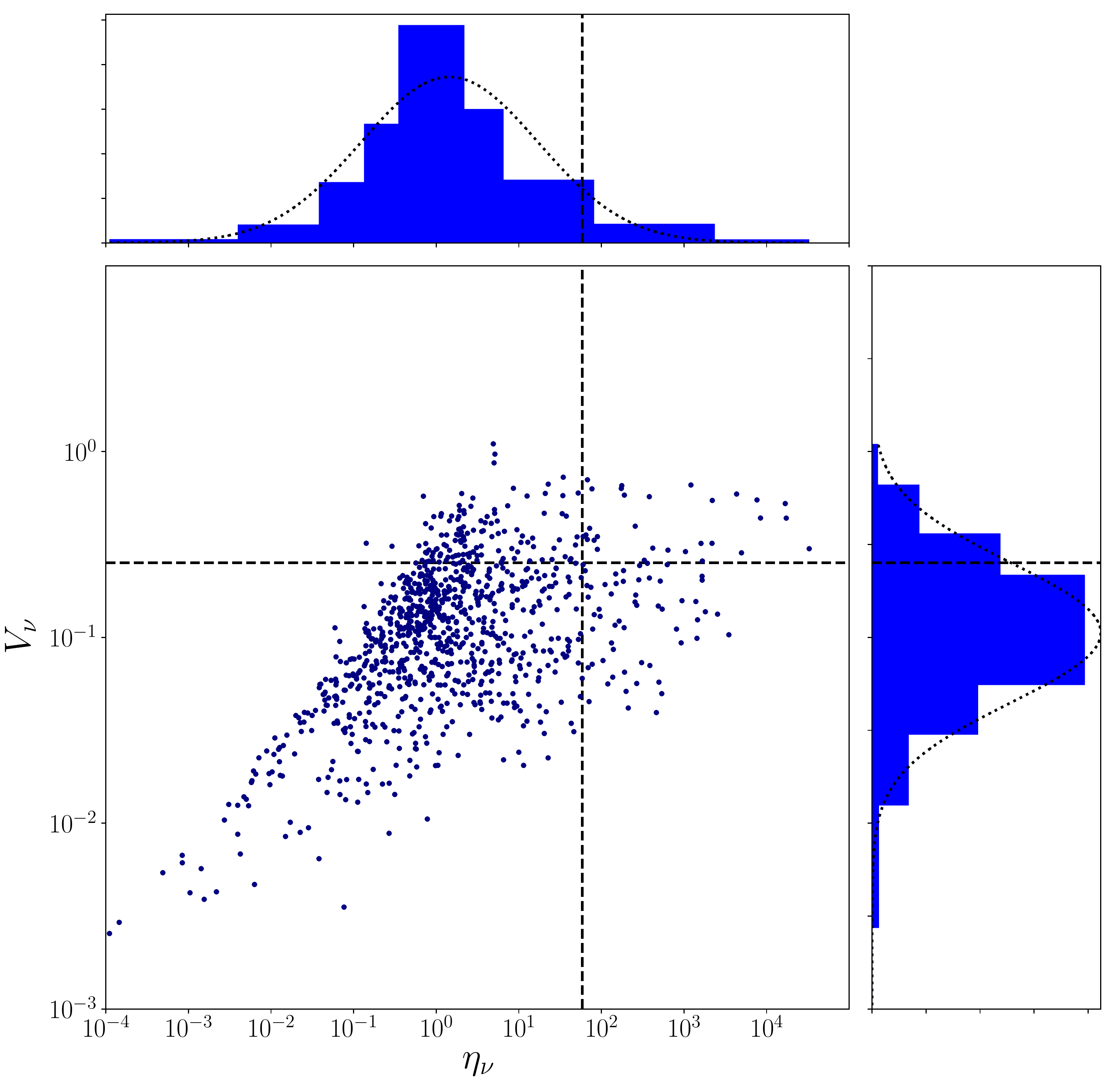}
    \caption{Histogram of the 2-dimensional distribution of the variability statistics $\eta$ and $V$ calculated by TraP for each source with two or three flux density measurements in our three epochs of observation.}
    \label{fig:varstat}
\end{figure}
\end{document}